\documentstyle[12pt,twoside,fleqn,espcrc1,epsfig]{article}

\newcommand{\AmS}{{\protect\the\textfont2
  A\kern-.1667em\lower.5ex\hbox{M}\kern-.125emS}}

\title{First Hints of Jet Quenching at RHIC} 

\author{A. Drees, SUNY at Stony Brook}       
\begin{document}

% typeset front matter
\maketitle

\begin{abstract}
At this conference first data from RHIC has been presented. Spectra of
charged hadrons and identified neutral pions obtained in central
collisions exhibit a depletion at large transverse momenta compared to
expectations deduced from $pp$ and $\bar{p}p$ data and  lower energy
heavy ion data. While spectra measured in peripheral collisions exhibit 
the expected power-law shape, spectra from central collisions are 
closer to exponential. In addition, a significant azimuthal anisotropy of high 
momentum charged particle production has been found. All observations are 
in qualitative agreement with theoretical predictions that quark matter
formed in heavy ion collisions quenches jet production. 
\end{abstract}

\section{Introduction}

Parton-parton scattering with large momentum transfer, so-called hard
scattering processes, provide unique probes for studing heavy ion
collisions. The reason is simple: hard scattering occurs early in the
collision, well before quark matter is expected to form. The
scattered partons will sense the full space-time evolution of the
collision volume and thus probe the later formed hot and dense phase. 

Scattering with large momentum transfer either results in the production 
of high mass particles, like charm quarks of which a small faction bind in 
$J/\psi$ charmonium states, or in high momentum quarks or gluons 
which fragment into jets of hadrons. If the scattered partons penetrate 
quark matter significant modifications of $J/\psi$ production and jet
fragmentation are expected. In fact, the discovery of charmonium
suppression \cite{na50} is one of the corner stones of the argument that 
quark matter has been formed already at CERN energies \cite{carlos}. 
Jet production should also be altered significantly. High momentum
partons should lose a significant fraction of their energy by 
gluon bremsstrahlung effectively suppressing jet production. This
phenomenon, commonly refered to as ``jet quenching'', was predicted a
decade ago \cite{guy1}. 

Among the large number of produced particles present in the final state 
of a heavy ion collision jets can not be reconstructed directly.
However, one of the jet fragments will always carry a major fraction of
the jet momentum. These so-called leading particles manifest themselves
in a power-law shape of the transverese momentum distribution. If
jets are quenched a depletion of the high momentum tail
of the spectra is expected. In addition, due to the binary nature of
hard scattering processes, most jets are produced in pairs, thus
azimuthal correlations between high $p_\perp$ particles might also 
serve as an experimental observable.  

At CERN SPS energies measurements of high $p_\perp$ particle production 
where performed \cite{wa98,ceres}. A conclusive interpretation failed 
so far, mostly because the data can be explained by hard scattering 
\cite{wang-0} as well as transverse flow \cite{heinz}. In contrast, first
data from RHIC show characteristic features consistent with the 
anticipated jet quenching. This talk summarizes the experimental 
evidence shown at QM2001. In the next two sections data from 
nulceon-nucleon collisions are evaluated and used to establish 
a reference to which RHIC data can be compared. 
Section 4 compares inclusive charged particle production to the reference. 
Additional information on identified particle production,
and its implicitions are discussed in section 5.  The final section
summarizes the talk. 

\section{Elementary reactions}

\begin{figure}[htb]
\vspace*{-1.2cm} 
  \begin{minipage}{0.49\linewidth}
  \begin{flushleft}
     	\epsfig{file=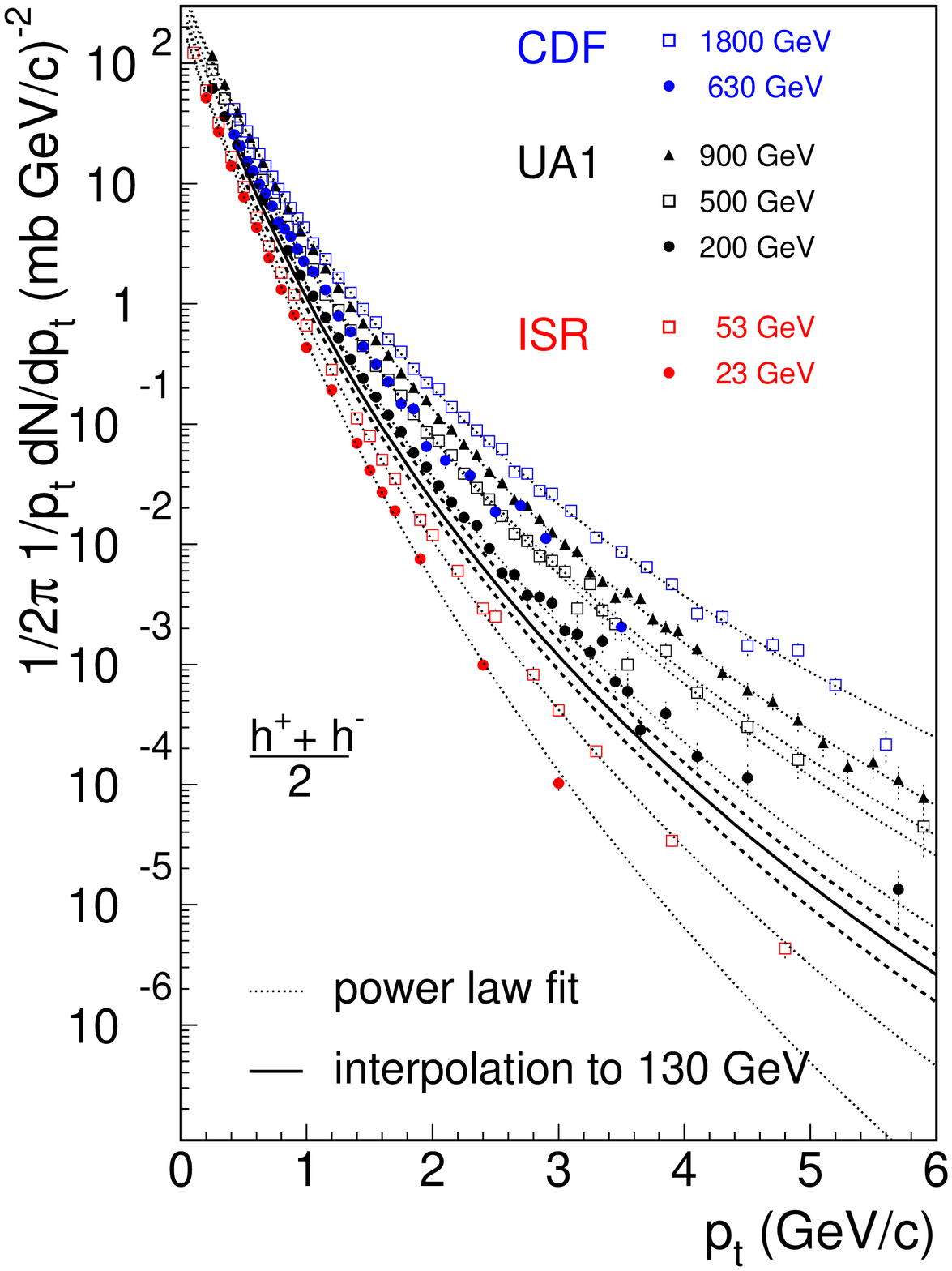,width=1.05\linewidth}
  \end{flushleft}
  \end{minipage}
  \begin{minipage}{0.49\linewidth}
  \begin{flushright}
     	\epsfig{file=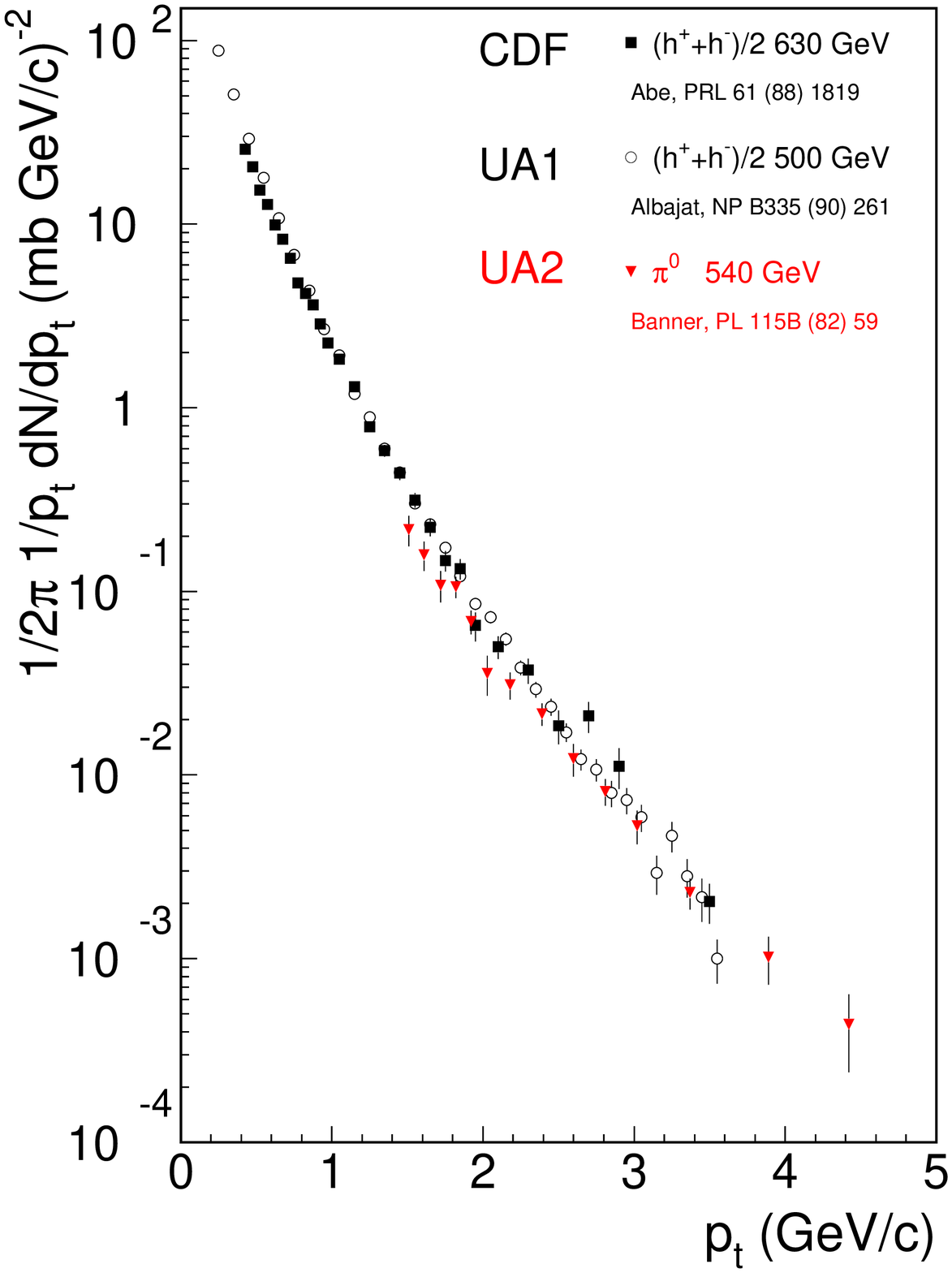,width=1.05\linewidth}
  \end{flushright}
  \end{minipage}

\vspace*{-2.4cm}
  \begin{flushleft}
  \begin{minipage}{0.47\linewidth}
     	\caption{{\label{fig:ppinter}} \footnotesize
        Inclusive charged particle production for $pp$ and $\bar{p}p$
        collisions \cite{alper,ua1,cdf}. The ISR data for 
        $\pi, K, p,$ and $\bar{p}$ were added to 
	obtain the charged spectrum. All data were fitted 
        with a power-law function (dotted lines) and interpolated to 
        $\sqrt{s} =130$ GeV (thick line), systematic errors 
        of the interpolation are shown as dashed lines.}
  \end{minipage}
  \end{flushleft}
  \vspace*{-4.85cm}
  \begin{flushright}
  \begin{minipage}{0.47\linewidth}
     	\caption{{\label{fig:ua2}} \footnotesize
     	Charged particle and neutral pion data from different experiments 
        \cite{ua1,ua2,cdf} at  $\sqrt{s}$ from 500 to 630 GeV. The 
	consistency of the data is used to estimate the systematic errors 
	on the absolute normalization of individual experiments. }
  \end{minipage}
  \end{flushright}
\vspace*{-0.4cm} 
\end{figure}

Searching for new phenomena in heavy ion collisions requires a detailed 
understanding of elementary nucleon-nucleon collisions. 
Data on high momentum particle production from $pp$ or $\bar{p}p$ collisions
exists for various beam energies, but not for the initial RHIC energy of
$\sqrt{s}_{nn}=130$ GeV. Fig.\ref{fig:ppinter} shows inclusive charged
particle $p_\perp$ spectra obtained from $pp$ collisions at the ISR
\cite{alper}, and from $\bar{p}p$ collisions by UA1 \cite{ua1} at CERN,
and by CDF at FNAL \cite{cdf}. With increasing beam energy high
$p_\perp$ particle production is enhanced,  reflecting
the increase of the jet production cross-section. 
    
To generate a reference $p_\perp$ distribution for
the initial RHIC energy $\sqrt{s}_{nn}=130$ GeV, the data shown in
Fig.\ref{fig:ppinter} were fitted with the empirical functional form
$d^2\sigma/dp_\perp^2 = A/(p_{0}+p_\perp)^n$. The cross sections were
then interpolated to 130 GeV at several fixed $p_\perp$. Finally, the
interpolated cross sections were fitted with the identical power-law
function to obtain a smooth reference distribution. The fit parameters
obtained are $A = 2.75\ 10^5$,  $p_{0} = 1.71$, and $n =12.42$. The fits
and the interpolation are shown in the figure. The dashed lines below
and above the interpolation indicate the systematic uncertainty which
results from (i) uncertainties in the interpolation procedure which
increases with $p_\perp$ and more importantly (ii) systematic
discrepancies of data sets take by different experiments. The latter
error was estimated to by about 20\% by comparing different data sets at
similar $\sqrt{s}$. An example is given Fig.~\ref{fig:ua2}
\cite{ua1,cdf,ua2}.

The same magnitude of systematic errors on the nucleon-nucleon reference 
is estimated by comparing to other extrapolations done by the STAR 
collaboration \cite{tu,star} and X.N. Wang \cite{wang-1} as well as 
comparing it to theoretical calculations of the cross section 
\cite{ivan,wang-2}.

\section{Extrapolating to $AA$ and the Nuclear Modification Factor}

To extrapolate from $pp$ to $AA$ collisions non-trivial assumptions need to
be made. Lets first assume all high $p_\perp$ particle production
results from binary hard collisions. In the absence of nuclear effects
the cross section should then simply scale by the atomic number squared
$A^2$. For a specific impact parameter selection corresponding to a
fraction $f$ of the inelastic nucleus-nucleus cross section 
$\sigma_{AA}^{inel}$, the average  number of binary collisions 
of the event sample $\langle N_{binary} \rangle$ can be calculated
from the nuclear overlap integral and the inelastic 
nucleon-nucleon cross section $\sigma_{nn}^{inel}$ \cite{eskola}:  

\begin{equation}
\langle N_{binary} \rangle 
= \sigma_{nn}^{inel} \ 
  \frac{\int_{b_1}^{b_2}{d^2b\ T_{AA}(b)}}{\int_{b_1}^{b_2}{d^2b}} 
= \sigma_{nn}^{inel} \ 
  \frac{\int_{b_1}^{b_2}{d^2bT_{AA}(b)}}{f \sigma_{AA}^{inel}} 
= \sigma_{nn}^{inel} \langle T_{AA} \rangle
\label{equ:scaling}
\end{equation}

In the following values of $\sigma_{nn}^{inel}=42\ mb$ and
$\sigma_{AA}^{inel}=7200\ mb$ are used consistently.  
Experimentally, the number of binary collisions can be evaluated  
from measured global observables like the forward energy
or the event multiplicity by a Glauber model approach (e.g. in \cite{mult}). 
To search for deviations from the $nn$-reference one may divide the 
$p_\perp$ spectra from heavy ion collisions by the scaled $nn$-reference. 
This ratio, denoted as $R_{AA}$, will be referred to as the
{\it nuclear modification factor} following the suggestion from \cite{wang-1}. 

\begin{equation}
R_{AA}(p_\perp) =\frac{d\sigma_{AA}/dyd^2p_\perp}{\langle N_{binary}\rangle  
d\sigma_{pp}/dyd^2p_\perp}
\label{equ:raa}
\end{equation}

If simple scaling with the number of binary collisions holds true 
the nuclear modification factor $R_{AA}$ should be unity. At low
momenta, say below 1 GeV/c, $R_{AA}$ must be smaller than one 
since in this $p_\perp$ region the cross section is expected to scale
with the number of participating nucleon rather than with the number of
binary collisions. Thus scaling with $\langle N_{binary} \rangle$ is
expected only at high $p_\perp$, say above 2-3 GeV/c. 

\begin{figure}[htb]
\vspace*{-0.6cm}
  \begin{minipage}{.51\linewidth}
  \begin{flushleft}	
  \begin{center}
     	\epsfig{file=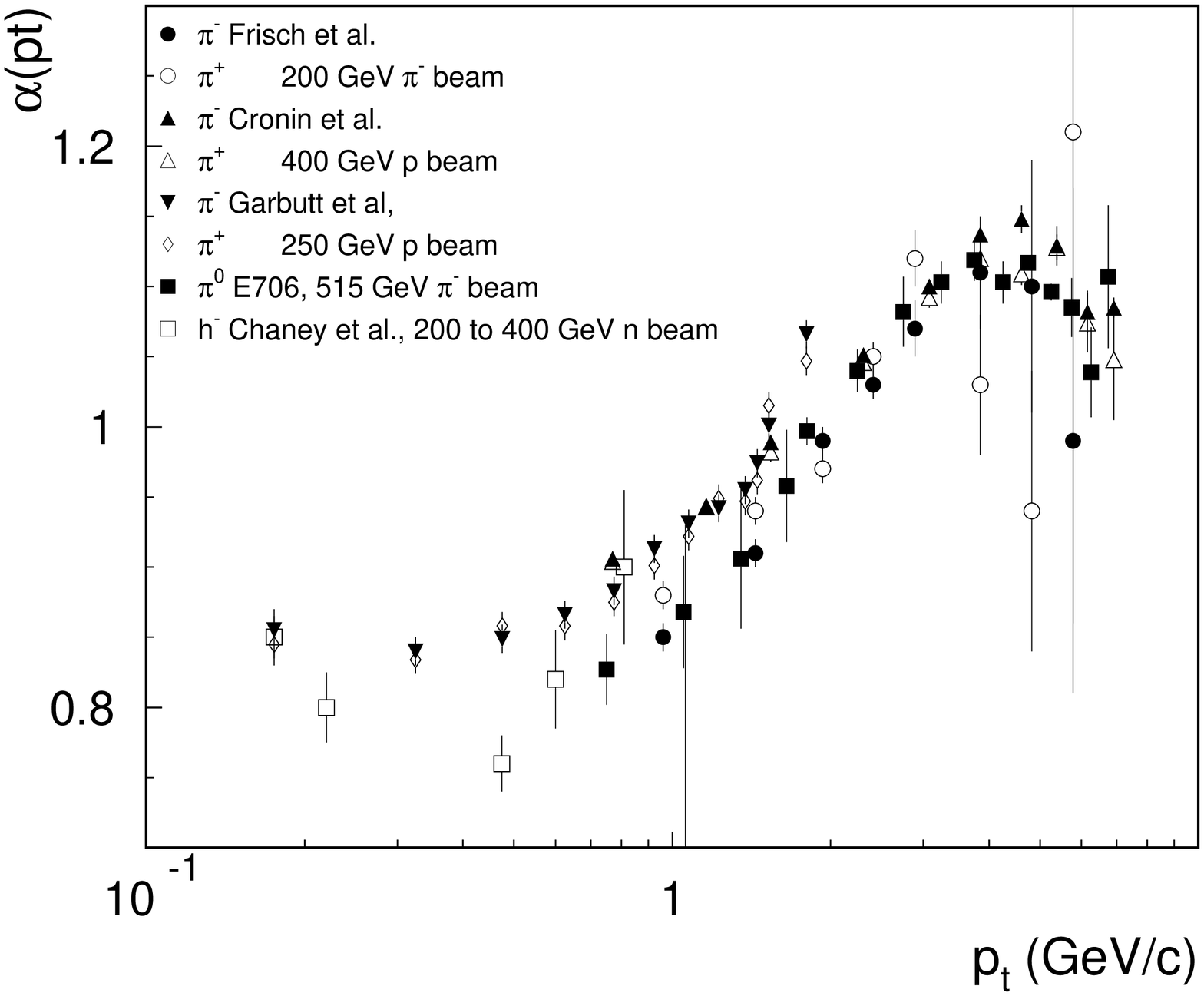,width=1.03\linewidth}
  \end{center}
  \end{flushleft}
  \end{minipage}
  \begin{minipage}{.47\linewidth}
  \begin{flushright}	
  \begin{center}
     	\epsfig{file=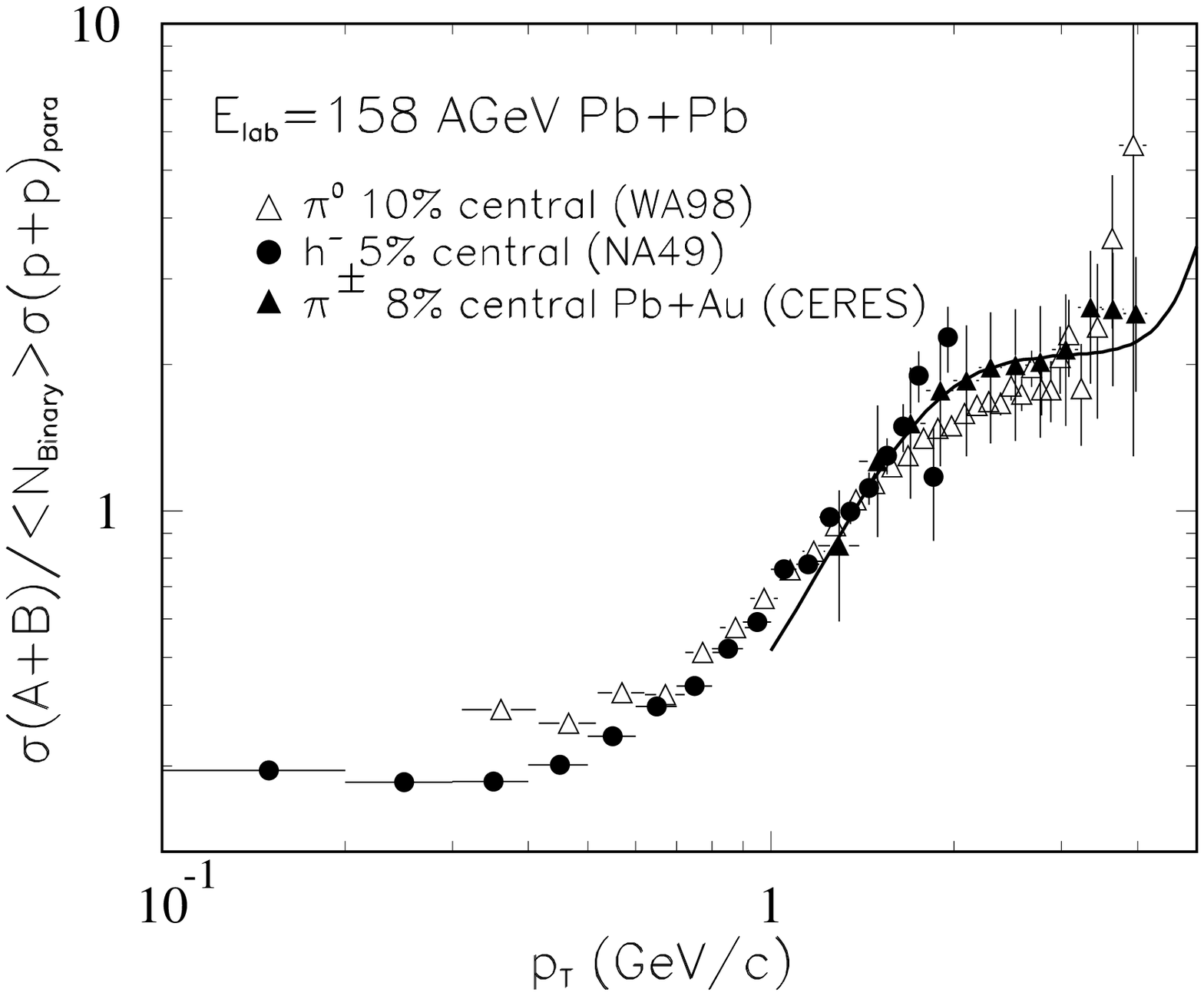,width=1.05\linewidth}
  \end{center}
  \end{flushright}
  \end{minipage}

\vspace*{-1.6cm}
  \begin{flushleft}	
  \begin{minipage}{.51\linewidth}
   \begin{center}
     	\caption{{\label{fig:cronin}} \footnotesize
     	Compilation of data for the Cronin exponent $\alpha(p_\perp)$ 
        \cite{cronin,list}. The excess above binary scaling, the 
        ``anomalous nuclear'' enhancement, sets in around 2 GeV/c when 
        $\alpha(p_\perp)$ increases above unity.}
  \end{center}
  \end{minipage}
  \end{flushleft}

\vspace*{-4.01cm}
  \begin{flushright}	
  \begin{minipage}{.47\linewidth}
  \begin{center}
     	\caption{{\label{fig:cern-nmf}} \footnotesize
     	 Nuclear modification factor for charged and pion data 
         obtained in 158 AGeV Pb-Pb and Pb-Au collisions at CERN 
         \cite{wa98,ceres,na49} (figure 5 from \cite{wang-1}). As shown in 
         \cite{wang-1} the nuclear modification factor is consistent 
         with the Cronin effect.}
  \end{center}
  \end{minipage}
  \end{flushright}
\vspace*{-1.7cm} 
\end{figure}

Below $p_\perp$ of 10 GeV/c the situation is complicated by known
nuclear effect. Already in the late 70's it was discovered by  
Cronin et al. \cite{cronin} that high $p_\perp$ particle production in
p-nucleus collisions is enhanced beyond simple binary
scaling. Traditionally this enhancement, now called Cronin effect, has
been parameterized as $\sigma_{pA} = A^{\alpha(p_\perp)}\ 
\sigma_{pp}$. Fig.\ref{fig:cronin} gives a compilation of the $p_\perp$ 
dependent exponent from different fixed target experiments \cite{cronin,list}. 
Though, so far there is no quantitative theoretical description of the
Cronin effect, it is commonly accepted that it originates from initial state
multiple scattering, i.e. multiple scattering of parton in the 
collision \cite{lev}. 

The Cronin effect has also been observed in heavy ion collisions
\cite{helios}. Data taken at CERN has been recently compiled and analyzed 
in terms of the nuclear modification factor \cite{wang-1}. 
Fig.\ref{fig:cern-nmf} reproduces a figure from \cite{wang-1} 
showing $R_{AA}$ for Pb-Pb
collisions. $R_{AA}$ continuously increases, crosses unity around 1.5
GeV/c and eventually saturates above 2.5 GeV at a value $R_{AA}\ \sim$ 2. 
At higher energies a similar behavior is expected, though one
might expect a reduction with increasing energy \cite{wang-1}.  

In summary, if previously observed phenomena are included, at RHIC
$R_{AA}$ should be above the naive binary scaling, which probably 
underestimates $R_{AA}$ and below the empirically observed value from 
CERN, which might overestimate $R_{AA}$. This conclusion implicitly assumes
that (i) the relative particle abundances -- in particular at high
 $p_\perp$ --  do not change going from nucleon-nucleon to
nucleus-nucleus extracted from RHIC data collisions, and that (ii) 
collective radial flow as well as (iii) shadowing 
do not alter the spectra at high $p_\perp$ significantly. 

\section{Inclusive charged particle data from RHIC}

\begin{figure}[htb]
\vspace*{-1.cm}
  \begin{minipage}{.44\linewidth}
  \begin{flushleft}	
    \begin{center}
     	\epsfig{file=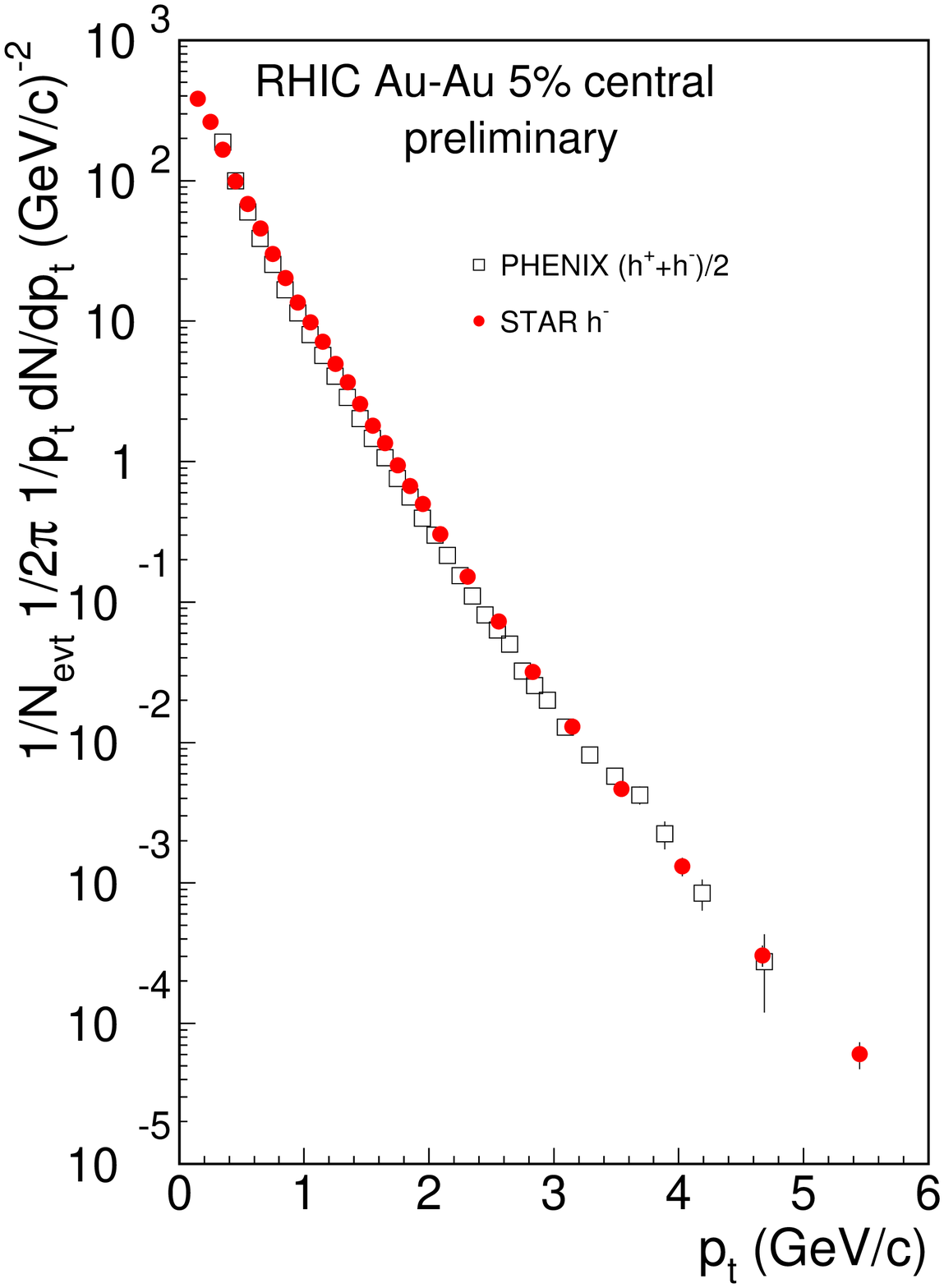,width=1.1\linewidth}
    \end{center}
  \end{flushleft}
  \end{minipage}
  \begin{minipage}{.54\linewidth}
  \begin{flushright}	
  \begin{center}
     	\epsfig{file=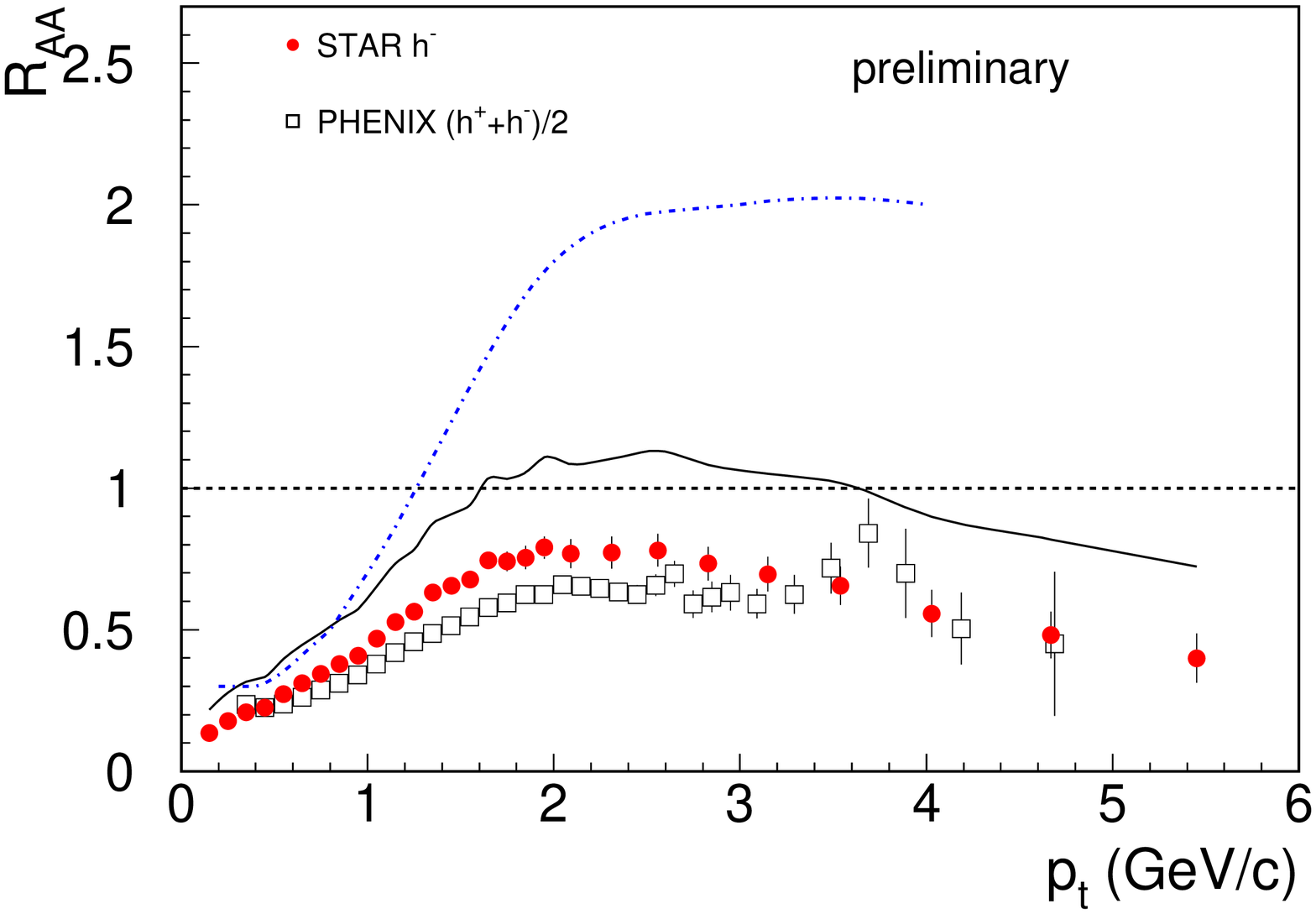,width=1.1\linewidth}
  \end{center}
  \end{flushright}	
  \end{minipage}

\vspace*{-2.4cm}

  \begin{flushleft}	
   \begin{minipage}{.44\linewidth}
   \begin{center}
     	\caption{{\label{fig:pt_compare}} \footnotesize
     	Comparison of data on negatively charged hadrons obtained by 
        STAR \cite{star} and charged hadrons from PHENIX \cite{federica} 
        for the most central 5\% 
        of the collisions. Both data sets are independently absolutely 
        normalized. }
    \end{center}
  \end{minipage}
  \end{flushleft}
\vspace*{-5.7cm}
  \begin{flushright}	
  \begin{minipage}{.52\linewidth}
  \begin{center}
     	\caption{{\label{fig:pt_nmf}} \footnotesize
     	Nuclear modification factor determined as ratio of the data from 
        figure \ref{fig:pt_compare} over the $nn$-reference described 
        in section 2, scaled by $\langle N_{binary} \rangle$ = 945 
        (see text for details). The thin line indicates the estimated 
         systematic uncertainty. Also given is the expectation for
         scaling with  $\langle N_{binary} \rangle$ (dashed line) and the 
         nuclear modification factor deduced from CERN data
         (dash-dotted line)}
  \end{center}
  \end{minipage}
  \end{flushright}	
\vspace*{-1.9cm}
\end{figure}

Fig.\ref{fig:pt_compare} compares the inclusive $p_\perp$ distribution of
negatively charged hadrons from STAR \cite{star} to the spectra of all
charged hadrons presented by PHENIX \cite{federica}. Both data sets are
obtained from the 5\%  most central Au-Au collisions at pseudo-rapidity zero. 
The data are consistent over the entire range observed out to 5 GeV/c. 
This is remarkable, given the preliminary nature of the data and
the known difficulties of absolute measurements. There are subtle
difference between the data sets which amount less than 40\% and are
largest around 2 GeV/c. This difference is used to estimate the
systematic error on the absolute normalization. In the following the
error is assumed to be $\pm$20\%, which is consistent with the 
errors quoted by the experiments. 

To compare to the $nn$-reference the nuclear modification factor is
calculated. The $nn$-reference was scaled by 
$\langle N_{binary} \rangle = 945\pm 140$ determined by PHENIX
\cite{mult}. STAR used a value of $\langle T_{AA}\rangle =
26 \ mb^{-1}$, corresponding to $\langle N_{binary} \rangle = 1092$
\cite{star} to scale their reference. The difference of both values 
is used as systematic uncertainty of 15\% on $\langle N_{binary}
\rangle$. This estimate is consistent with the $\sim$
15\% error given by PHENIX. 

The resulting nuclear modification 
factor $R_{AA}$ is shown in Fig.\ref{fig:pt_nmf}. On a linear scale the 
subtle differences between the data sets become more apparent. 
The systematic errors on $R_{AA}$ are indicated by the full line, they
results from (i) the error on the $nn$-reference ($\sim$20\% but 
$p_\perp$ dependent), (ii) a 20\% error on the normalization, and (iii) 
a 15\% error on $\langle N_{binary} \rangle$. The errors quoted here are
somewhat larger than those evaluated by PHENIX and STAR, the reason is
two fold: (i) the errors take into account the differences between the 
data and (ii) the error on the $nn$-reference was estimated more
conservatively.   

Initially, $R_{AA}$ increases up to a $p_\perp$ of about 2 GeV/c where 
it saturates at a value of 0.6 to 0.8. At high $p_\perp$ the nuclear 
modification factor $R_{AA}$  seems to decrease again. In the region 
from 2 to 3 GeV $R_{AA}$ might be consistent with unity within the 
rather large systematic uncertainties. However, it is not converging 
towards one at higher $p_\perp$, as suggested by simple binary scaling. 
The figure also indicates $R_{AA}$ observed at the CERN-SPS 
(dash-dotted line). Obviously, $R_{AA}$ at RHIC is significantly 
below the value found for CERN SPS data.

\begin{figure}[htb]
\vspace*{-1.1cm}
  \begin{minipage}{1.\linewidth}
  \begin{center}
      	\epsfig{file=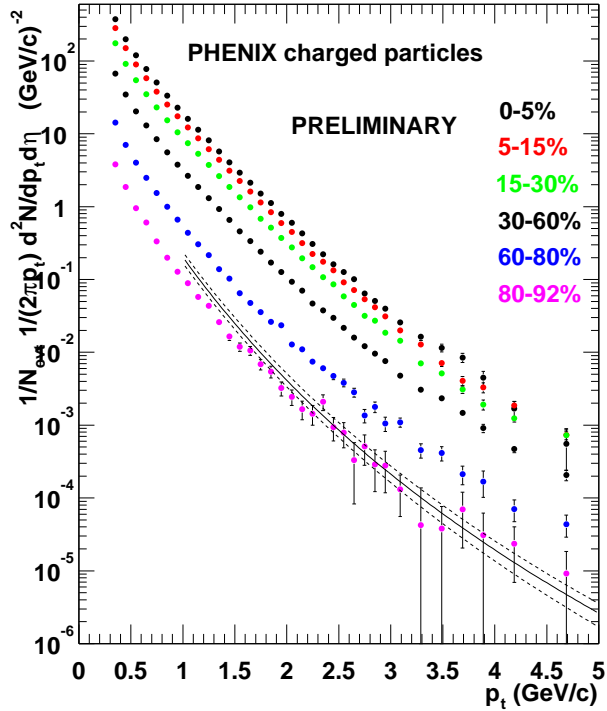,width=.54\linewidth}
\vspace*{-1.cm}
     	\caption{{\label{fig:phenix_charged}} \footnotesize
     	Inclusive charged particle production measured by PHENIX 
        \cite{federica}
        for different centrality selections (specified in the figure). 
        All data sets are absolutely normalized. The lowest data set 
        represents the most peripheral selection and 
        the highest data set the most central selection. }
  \end{center}
  \end{minipage}
\vspace*{-.8cm}
\end{figure}

\begin{figure}[htb]
\vspace*{-.5cm}
  \begin{minipage}{0.48\linewidth}
  \begin{center}
     	\epsfig{file=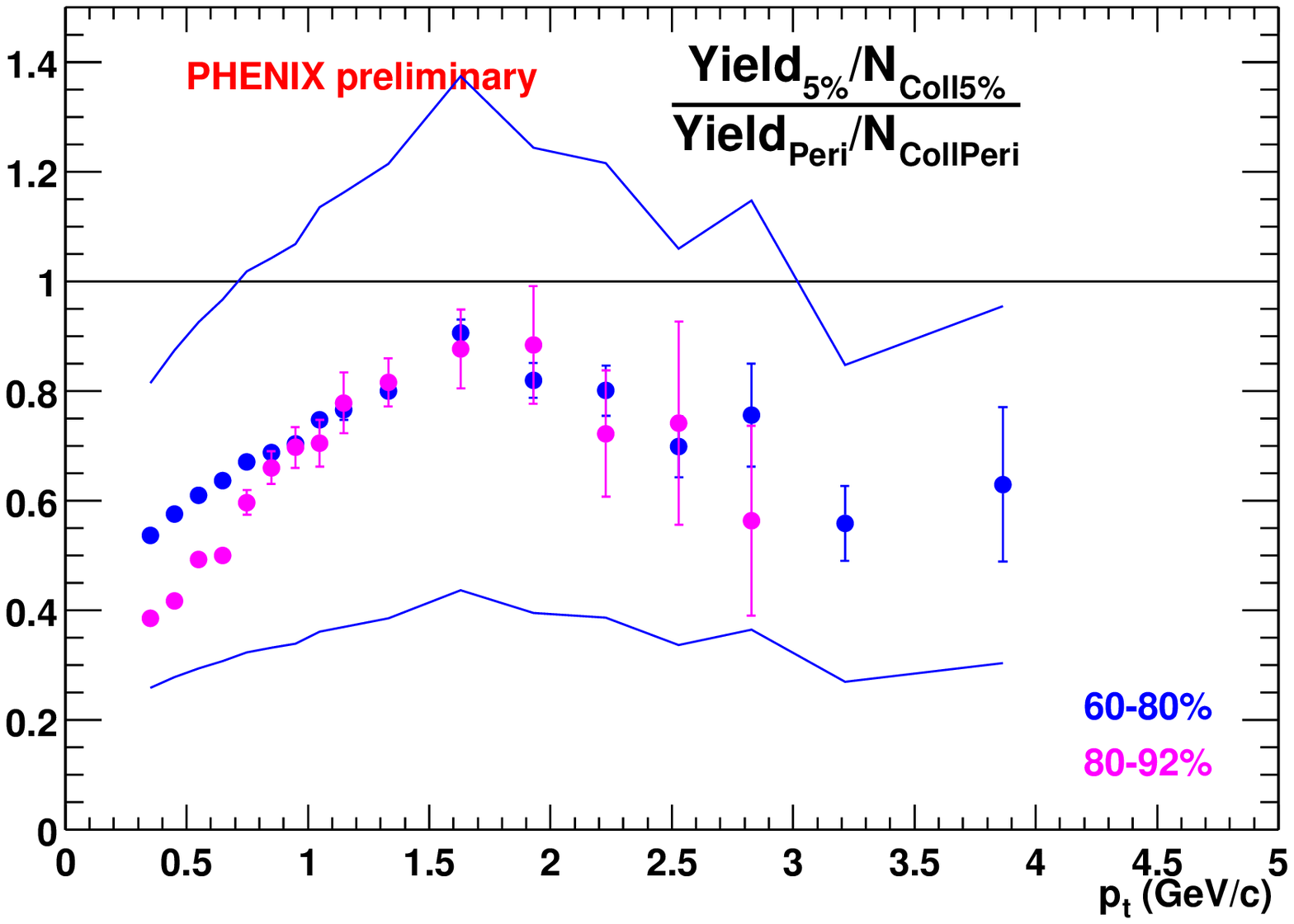,width=1.05\linewidth}
  \end{center}
  \end{minipage}
  \begin{minipage}{0.48\linewidth}
  \begin{center}
     	\epsfig{file=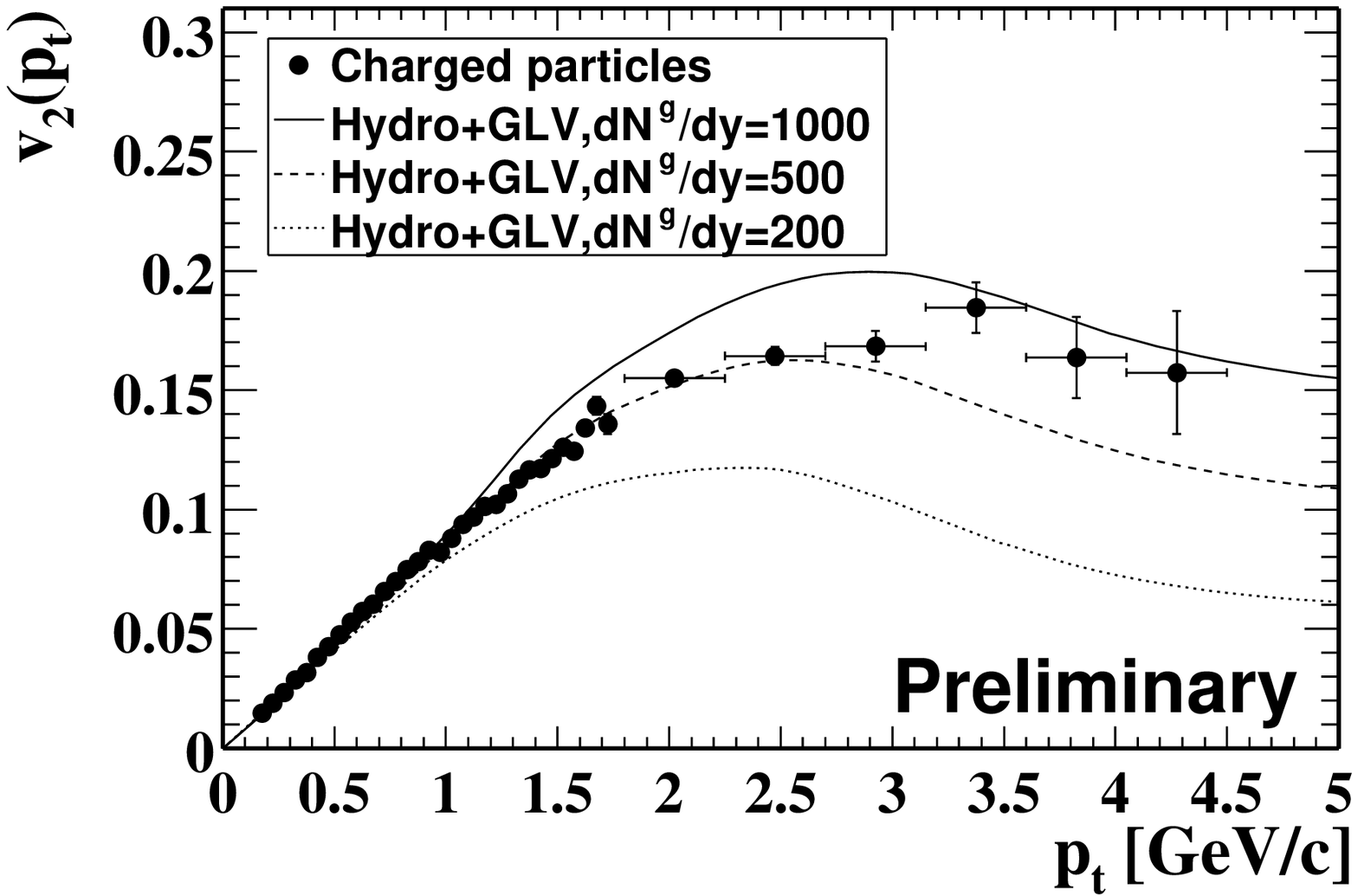,width=1.15\linewidth}
  \end{center}
  \end{minipage}
\vspace*{-1.1cm}
  \begin{flushleft}
  \begin{minipage}{0.48\linewidth}
     	\caption{{\label{fig:ratio_cp}} \footnotesize
     	Ratio of central to peripheral data on charged 
	particle production (as shown in Fig.\ref{fig:phenix_charged}). The 
        lines indicate the systematic error band, which results mostly from 
        the uncertainty in the number of  binary collisions for the 
        peripheral event selections $\langle N_{binary} \rangle$.
	}
  \end{minipage}

  \end{flushleft}
\vspace*{-4.35cm}
  \begin{flushright}
  \begin{minipage}{0.48\linewidth}
     	\caption{{\label{fig:v2}} \footnotesize
     	 STAR data  \cite{snelling} on azimuthal anisotropy 
         -- $v_2$ -- measured versus 
         the reaction plane as a function of $p_\perp$. Also shown are 
         recent model calculations invoking a quark matter phase \cite{GLV}. 
         The three different lines indicate different initial gluon densities.}
  \end{minipage}
  \end{flushright}
\vspace*{-1.cm}
\end{figure}

PHENIX has also presented a systematic survey of different centrality
selections. The data are shown in Fig.\ref{fig:phenix_charged}. The
spectra for the most peripheral collisions exhibit a pronounced
power-law shape which seems to vanish for the most central
selection. The significant difference between central and peripheral
collisions is more clearly visible in the ratio, which
is depicted in Fig.\ref{fig:ratio_cp}. 
To compare to the binary scaling assumption and to $R_{AA}$ the cross
sections were normalized by the number of binary collisions for the
specific centrality selection. The large uncertainty in 
$\langle N_{binary} \rangle$ for the peripheral data sets ($\sim 60\%$)
is reflected by the error band also shown in the figure; this ia an
uncertainty on the
absolute scale of the ratio but not on its $p_\perp$ dependence.

Both ratios, central 5\% to 80-92\% and to 60-80\% show the same
trend: a rise up to $p_\perp$ of 1.5-2 GeV/c followed by a decrease at
higher momenta. Such a behavior is qualitatively in agreement with 
a power-law spectrum becoming more exponential with increasing
centrality. Within the systematic errors the ratio could be close to
one, but as in Fig.\ref{fig:pt_nmf} the ratio does not converge 
towards unity at large $p_\perp$ as suggested by the binary 
scaling scenario. If one assumes 
that peripheral collisions resemble nucleon-nucleon interactions one can 
interpret the ratio shown in Fig.\ref{fig:ratio_cp} as an alternative
measure of the $R_{AA}$. The ratio derived here and  $R_{AA}$ shown in 
Fig.\ref{fig:pt_nmf} agree reasonably well. Although the systematic errors 
have similar size they have very different origin.

That something interesting is happening at high $p_\perp$ is
corroborated by data on azimuthal angular correlations 
presented by STAR \cite{snelling}. The data is presented in Fig.\ref{fig:v2} 
in terms of the $p_\perp$ dependence of $v_2$. Here $v_2$ is the second
Fourier coefficient of the azimuthal track density distribution measured
with respect to the reaction plane. The coefficient $v_2$ measures an
anisotropy of particle production relative to the reaction plane, 
which is typically interpreted as elliptic flow. At low
$p_\perp$ $v_2$ continuously increases at a rate predicted by 
model calculations based on hydrodynamics \cite{pasi}. Around 2 GeV/c 
the pattern changes and $v_2$ seems to saturate. At high $p_\perp$, say
above 2 GeV/c, particle production can no longer be described by
hydrodynamics and consequently the anisotropy is expected to vanish. 
Interestingly, the saturation of $v_2$ and $R_{AA}$ occur at 
similar $p_\perp$ suggesting a possible connection of both phenomena. 

Angular correlations between jets should not contribute to $v_2$
because jet production is uncorrelated to the reaction plane. 
If, however, jets loose significant energy in the dense medium an
anisotropy will be introduced by the geometry of the reaction volume. In
transverse direction, perpendicularly to the 
beam axis, the reaction volume will be almond shaped for non-zero impact
parameter. Thus, the average path length of a jet in the dense
medium will be smaller in plane than out of plane. It is plausible
that jets produced out of plane will loose more energy than 
those produced in plane, effectively leading to an anisotropy 
with respect to the reaction plane at medium and high momenta. 

The figure compares the data to a recent theoretical calculation
\cite{GLV} which combines a hydrodynamic calculation with perterbative
QCD, and modeles jet production quenched by energy loss. The different curves 
correspond to different initial gluon densities; for a density of 500
the model simultaneously describes also the $p_\perp$ spectra reasonably
well.

\begin{figure}[htb]
\vspace*{-0.8cm} 
  \begin{minipage}{0.47\linewidth}
  \begin{flushleft}
     	\epsfig{file=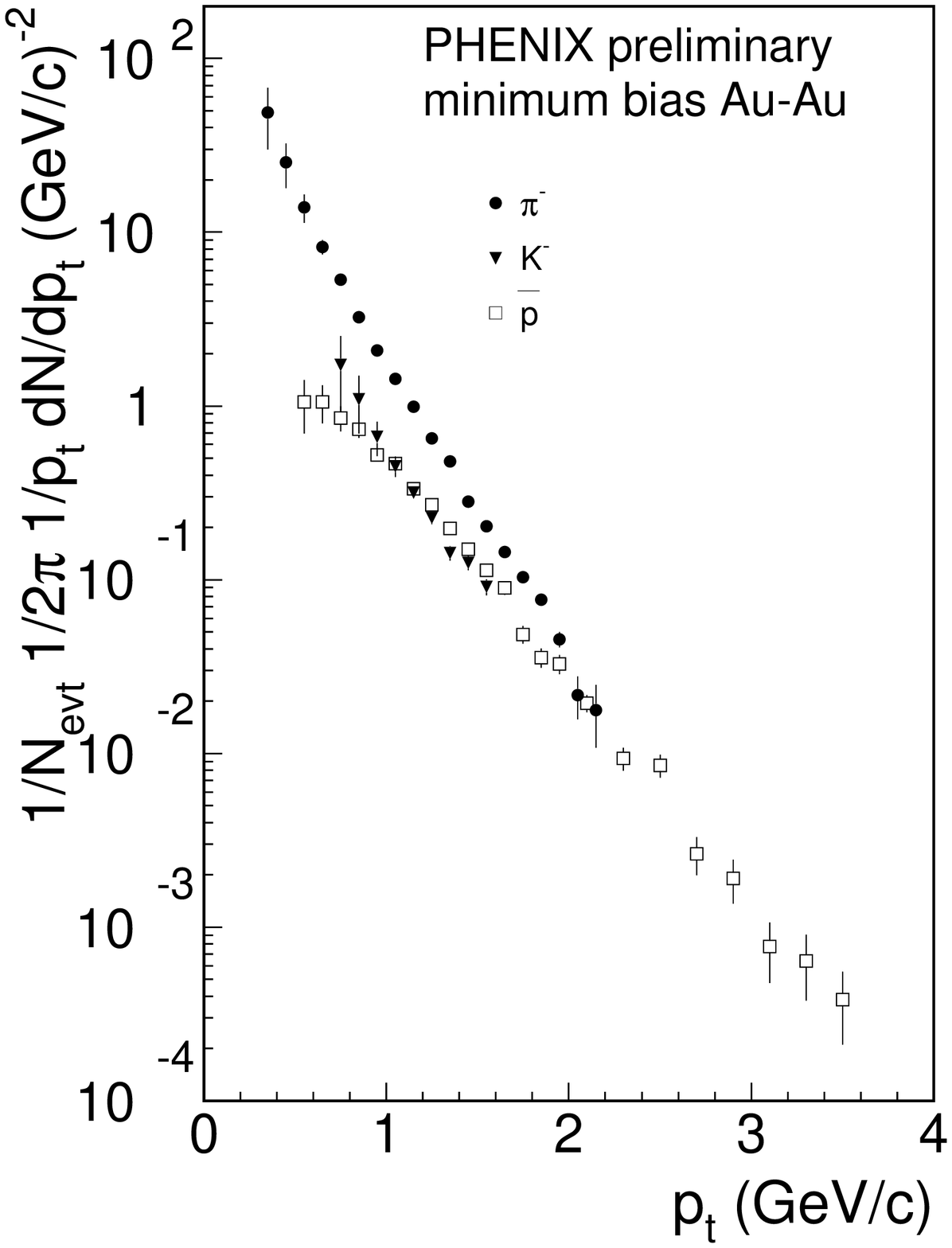,width=1.\linewidth}
  \end{flushleft}
  \end{minipage}
  \begin{minipage}{0.47\linewidth}
  \begin{flushright}
     	\epsfig{file=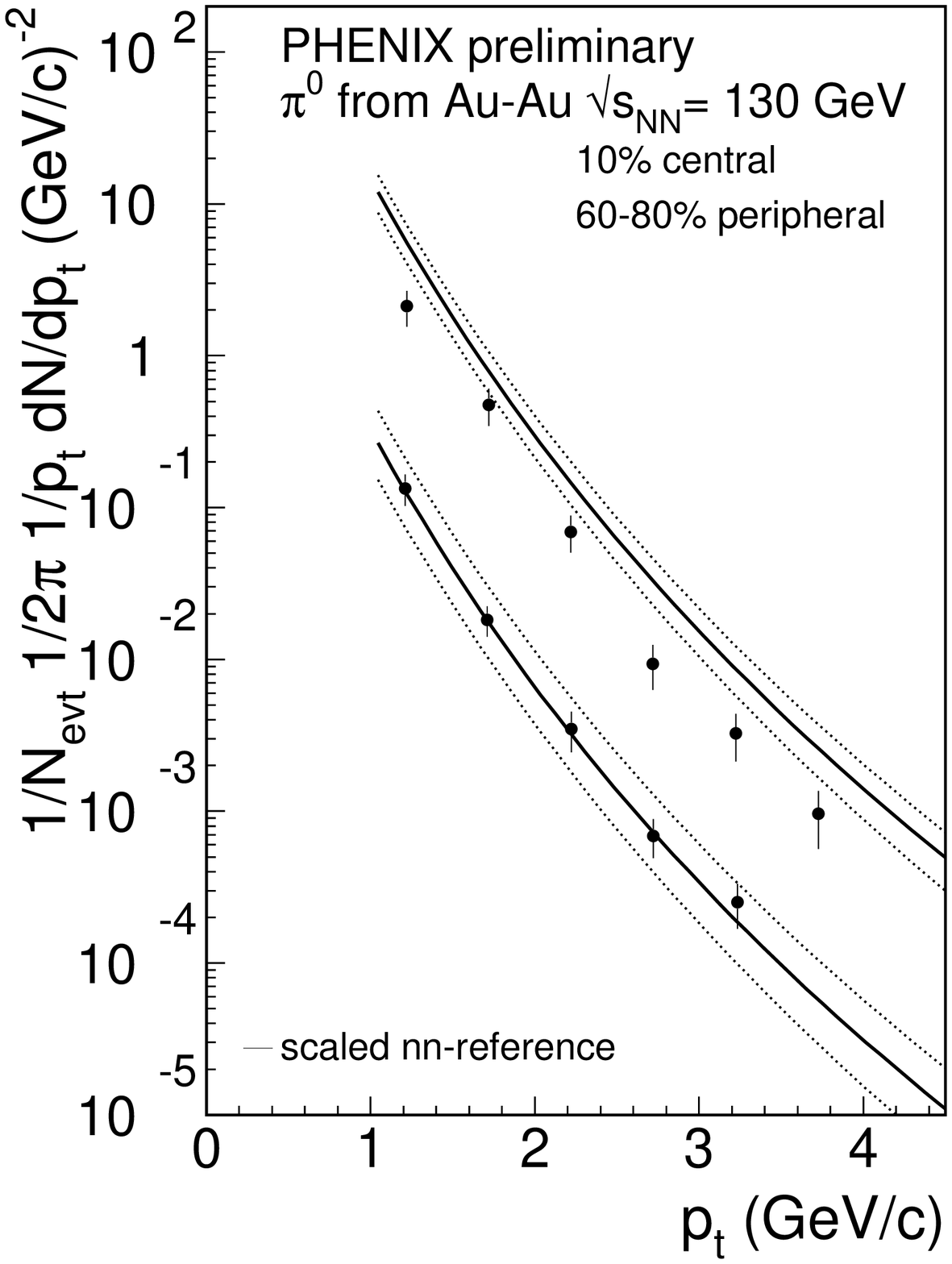,width=1.\linewidth}
  \end{flushright}
  \end{minipage}

\vspace*{-2.4cm}
  \begin{flushleft}
  \begin{minipage}{0.47\linewidth}
     	\caption{ {\label{fig:hadrons}} \footnotesize
        Inclusive $\pi^-$, $K^-$, and $\bar{p}$ production from minimum
        bias Au-Au collisions presented by PHENIX \cite{julia}. All three 
        distributions are absolutely normalized. }
  \end{minipage}
  \end{flushleft}
\vspace*{-3.55cm}
  \begin{flushright}
  \begin{minipage}{0.47\linewidth}
      	\caption{{\label{fig:pi0}} \footnotesize
     	 Neutral pion transverse momentum distribution measured by PHENIX 
         \cite{gabor}. The data is compared to the appropriately scaled 
         nucleon-nucleon reference for pion production 
         (systematic errors given as dashed lines).} 
  \end{minipage}
  \end{flushright}
\vspace*{-1.4cm}
\end{figure}

\section{Identified pion data from RHIC}

Additional information on high $p_\perp$ production of neutral pion, proton 
and anti-proton is available from PHENIX \cite{gabor,julia,ohnishi}. These 
first data show that in Au-Au collisions a much smaller fraction of the 
high $p_\perp$ charged particles are pions than in $pp$ collisions. 
Minimum bias data for $\pi^-,\ K^-$ and $\bar{p}$ is presented in  
Fig.\ref{fig:hadrons}. Above 2.0 GeV/c the ratio $\bar{p}/\pi^-$ (as
well as the $p/\pi^+$ ratio) approaches
unity, significantly larger than the value of $\sim 0.2$ 
observed in $pp$ collisions at the ISR 
\footnote{Value determined from data in \cite{alper}}. 
In addition, the spectra suggest that $\bar{p}/\pi^-$ continues to
increase towards higher $p_\perp$, quite different from $m_\perp$ 
scaling found in $pp$ collisions. Large collective radial flow in heavy  
ion collision might offer a possible explanation \cite{nu,teaney}. 

Independent of the interpretation, at high $p_\perp$ the particle
compositions in Au-Au at RHIC deviates from $pp$ collision where
the ratio $charged/\pi$ was about $1.6$ above 1.5
GeV/c. From Fig.\ref{fig:hadrons} one deduces $charged/\pi$ of about 2  
at 1.5 GeV/c, increasing to $\sim 2.5$ at about 2.2 GeV/c. 
Thus $R_{AA}$ measured for charged particles should not be interpreted 
as nuclear modification factor for pions. $R_{AA}$ underestimates the
depletion of pion production by approximately a factor $\sim$1.5. 

\begin{figure}[htb]
\vspace*{-1.8cm} 
  \begin{minipage}{0.39\linewidth}
  \begin{flushleft}
     	\epsfig{file=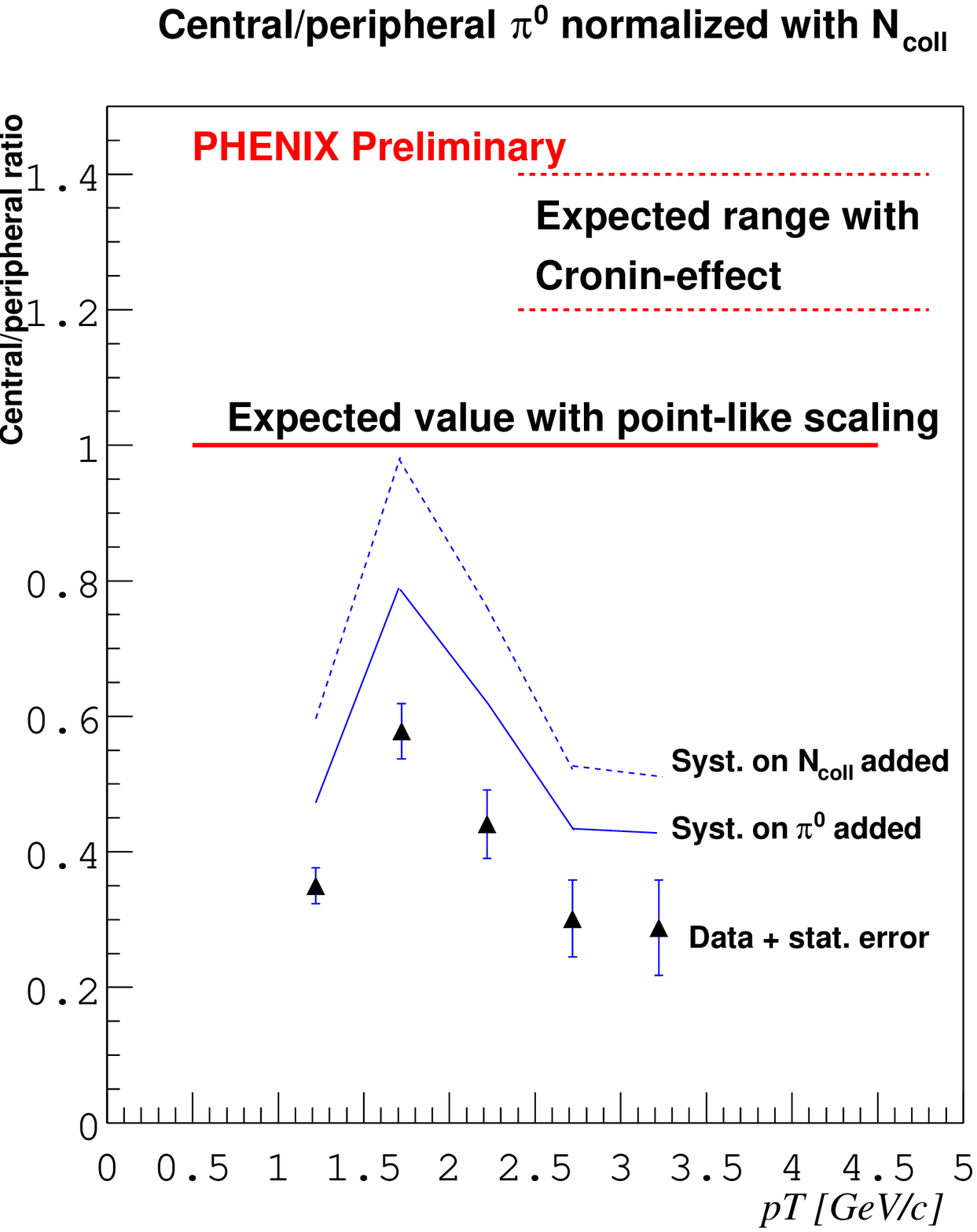,width=1.1\linewidth}
  \end{flushleft}
  \end{minipage}
  \begin{minipage}{0.59\linewidth}
  \begin{flushright}
   \vspace*{1.cm} 
     	\epsfig{file=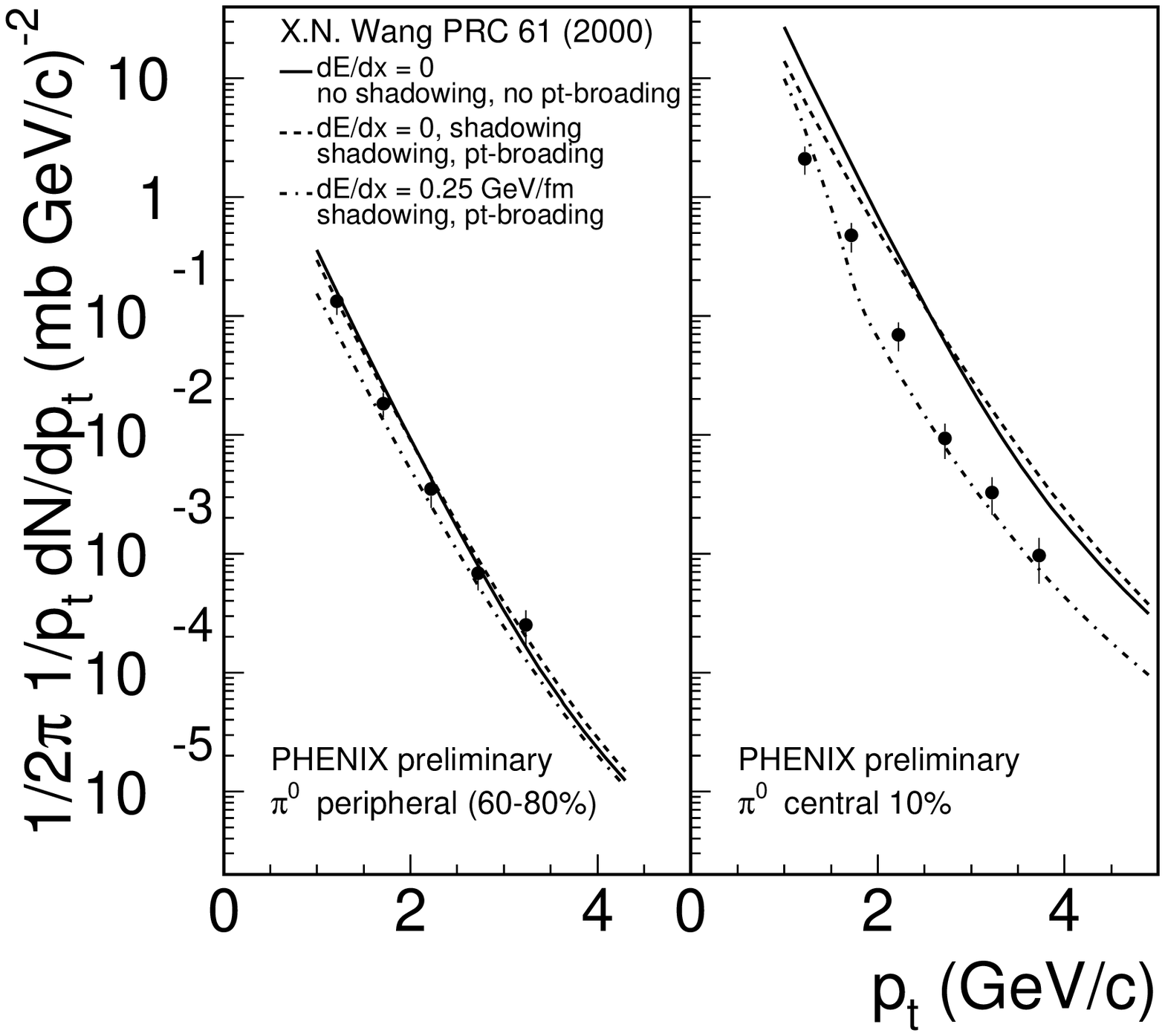,width=1.1\linewidth}
  \end{flushright}
  \end{minipage}

\vspace*{-2.6cm}
  \begin{flushleft}
  \begin{minipage}{0.47\linewidth}
     	\caption{ {\label{fig:ratio-pio-cp}} \footnotesize
        Ratio of central to peripheral neutral pion data. 
        The thin lines indicate the systematic uncertainty of the data.}
  \end{minipage}
  \end{flushleft}
\vspace*{-3.15cm}
  \begin{flushright}
  \begin{minipage}{0.47\linewidth}
      	\caption{{\label{fig:wang}} \footnotesize
     	 Comparison of neutral pion production to various model 
         calculations from \cite{wang-2}}
  \end{minipage}
  \end{flushright}
\vspace*{-0.7cm}
\end{figure}

The measurement of neutral pions by PHENIX \cite{gabor} reveals directly
the more substantial suppression of high $p_\perp$ pion production. Two  
data sets are given in Fig.\ref{fig:pi0}, a 10\% central and a
peripheral (60-80\%) event selection. The systematic uncertainty of the 
absolute normalization is included in the error bars. 

Both data sets are compared to the nucleon-nucleon reference. Here the 
reference, which was deduced for charged particles in section 2, was scaled 
down by the $charged/\pi$ ratio of 1.6 observed at the ISR. The systematic 
error bands shown in the figure contain the uncertainty of the 
reference (see section 2), a 10\% error on the $charged/\pi$ scale
factor, as well as 15\% and 60\% error on the number of binary collisions 
for central and peripheral data, respectively. While the peripheral data 
are well described by the reference distribution, the central data are 
significantly below. From the ratio of the data to the $nn$-reference 
one deduces
an average nuclear modification factor of $R_{AA} \sim 0.4$ for neutral
pions in central collisions. This value is smaller by a factor of 1.5 
compared to $R_{AA}$ obtained for charged particles, consistent with the
large $p$ and $\bar{p}$ yield at high $p_\perp$. For the peripheral data,
$R_{AA}$ is about unity reflecting the good agreement of the data and
the $nn$-reference. 

As alternative measure of $R_{AA}$ Fig.\ref{fig:ratio-pio-cp} gives 
the ratio of central to peripheral data. For all $p_\perp$ the ratio is 
below unity, indicating the substantial suppression of the yield in 
central collisions relative to the binary scaling assumption. Very 
similar to the charged particle data the ratio peaks between 1.5 and 
2 GeV/c and decreases towards higher $p_\perp$. 

Finally in figure \ref{fig:wang}, the neutral pion data are compared to 
theoretical model calculations 
by X.N. Wang \cite{wang-2}. Three different 
calculations are shown: (i) an estimate of the scaled $nn$-reference, 
(ii) a scaled reference modified by modeling the Cronin effect 
(here called $p_\perp$-broadening) and shadowing, as well as (iii) a 
calculation inferring ``jet quenching'' by introducing an average energy 
loss of 0.25 GeV/fm for the scattered partons. The peripheral data agrees 
reasonably well with all three scenarios while the central data are 
significantly below the calculations not including energy loss. 

\section{Conclusion and Outlook}

Only a few months after the end of its first Au-Au run, RHIC has produced 
a surprisingly rich sample of data, reaching out in $p_\perp$ well into 
a region where a significant contribution of hard scattering processes 
is expected. In central collisions a depletion of high $p_\perp$
particle production is discovered. This depletion 
is observed in charged particle production measured by STAR and PHENIX 
as well as in neutral pion data from PHENIX. The same depletion is found 
when comparing central to peripheral data. Additional information results 
from azimuthal anisotropy of particle production. All observations are 
consistent with a scenario in which quark matter formed during the collision 
suppresses jet production.  

Whether these first hints for ``jet quenching'' will hold true remains to 
be seen.  Drawing definite and quantitative conclusions from the present 
data certainly is premature in particular in view of the large systematic 
uncertainties. Besides better control of the absolute normalization it 
will be essential to measure the $nn$-reference in the same experiments. 
Data at higher $p_\perp$ will help to disentangle soft physics like radial 
flow from hard scattering.

At the end I would like to express my gratitude to many who have helped 
preparing my summary talk -- before and during this exciting conference. 
I specially thank T.Ullrich, P.Jacobs, and R.Snellings for providing the 
STAR data. X.N.~Wang generously provided calculations and figures 
for my talk. Also many colleges in PHENIX deserve credit, in 
particular G.David, F.Messer, and J. Velkovska. 

\def\IJMPA{{Int. J. Mod. Phys.}~{\bf A}}
\def\JPG{{J. Phys}~{\bf G}}
\def\NCA{Nuovo Cimento}
\def\NIM{Nucl. Instrum. Methods}
\def\NIMA{{Nucl. Instrum. Methods}~{\bf A}}
\def\NPA{{Nucl. Phys.}~{\bf A}}
\def\NPB{{Nucl. Phys.}~{\bf B}}
\def\PLB{{Phys. Lett.}~{\bf B}}
\def\PLC{Phys. Repts.\ }
\def\PRL{Phys. Rev. Lett.\ }
\def\PRD{{Phys. Rev.}~{\bf D}}
\def\PRC{{Phys. Rev.}~{\bf C}}
\def\ZPC{{Z. Phys.}~{\bf C}}
\def\EPJC{{Eur.Phys.J.}~{\bf C}}

\end{document}